\documentclass[aps,prl,notitlepage,nofootinbib,showpacs,twocolumn,floatfix]{revtex4-2}

\usepackage{graphicx}
\usepackage{amssymb}
\usepackage{amsmath}
\usepackage{bm}
\usepackage{latexsym}
\usepackage{epsfig}
\usepackage{psfrag}
\usepackage{ulem}
\usepackage{color}
\usepackage[dvipsnames]{xcolor}

\def\Mpl{M_{_{\rm Pl}}}
\def\beq{\begin{equation}}
\def\eeq{\end{equation}} 
\def\beqa{\begin{eqnarray}}
\def\eeqa{\end{eqnarray}}

\def\vka{{\bm k}_{1}}
\def\vkb{{\bm k}_{2}}
\def\vkc{{\bm k}_{3}}

\def\a1{\alpha_{_1}}
\def\b1{\beta_{_1}}
\def\de1{\delta_{_1}}
\def\g1{\gamma_{_1}}

\def\ns{n_{_{\rm S}}}

\def\ps{{\mathcal P}_{_{\rm S}}}

\def\fnl{f_{_{\rm NL}}}

\def\ki{k_{\rm i}}

\newcommand{\viz}{\textit{viz.~}}
\newcommand{\ie}{\textit{i.e.~}}

\begin{document}
\title{Unique contributions to the scalar bispectrum in `just enough inflation'} 
\author{H.~V.~Ragavendra$^\dag$, Debika Chowdhury$^\ddag$ and 
L.~Sriramkumar$^\dag$} 
\affiliation{$^\dag$Department of Physics, Indian Institute of Technology 
Madras, Chennai~600036, India\\
$^\ddag$Department of Theoretical Physics, Tata Institute of Fundamental 
Research, Mumbai~400005, India}

\begin{abstract}
A scalar field rolling down a potential with a large initial velocity 
results in inflation of a finite duration.
Such a scenario suppresses the scalar power on large scales improving 
the fit to the cosmological data. 
We find that the scenario leads to a hitherto unexplored situation 
wherein the boundary terms dominate the contributions to the scalar 
bispectrum over the bulk terms.
We show that the consistency relation governing the non-Gaussianity 
parameter~$\fnl$ is violated on large scales and that the contributions 
at the initial time can substantially enhance the value of~$\fnl$.
\end{abstract}
\maketitle


\noindent
\underline{\it Suppressing the power on large scales:}\/~It is well known
that a featureless and nearly scale invariant primordial spectrum, as 
is generated in slow roll models of inflation, is remarkably consistent 
with the observations of the anisotropies in the Cosmic Microwave 
Background~(CMB) (for the most recent constraints from Planck, see
Ref.~\cite{Akrami:2018odb}).
However, intriguingly, it has been repeatedly noticed that suppressing 
the primordial scalar power on large scales roughly corresponding to
the Hubble radius today improves the fit to the CMB data at the lower 
multipoles~\cite{Bridle:2003sa,Shafieloo:2003gf,Hunt:2004vt,Hunt:2007dn,
Hazra:2013nca}.
There has been a constant effort to construct models of inflation that 
naturally result in lower power on the largest observable
scales (see, for example, Refs.~\cite{Cline:2003ve,Contaldi:2003zv,
Powell:2006yg,Nicholson:2007by,Jain:2008dw,Jain:2009pm,Hazra:2014jka,
Hazra:2014goa}). 

\par

In the standard slow roll models of inflation, the scalar fields are 
assumed to start on the inflationary attractor, and they evolve along 
the attractor until the end of inflation.
A model that has drawn recent attention in the context of suppressing power
on large scales involves a scalar field which begins its journey down the 
inflationary potential with the largest initial velocity possible (for the 
original discussion, see Ref.~\cite{Contaldi:2003zv}; for recent discussions, 
see Refs.~\cite{Ramirez:2011kk,Ramirez:2012gt,Handley:2014bqa,Hergt:2018crm,
Hergt:2018ksk}).
In fact, in such a situation, inflation begins only after about an e-fold
or two, when the friction arising due to the expansion of the universe 
has reduced the velocity of the field adequately.
Thereafter, the field rolls slowly down the inflationary potential and, as
usual, inflation is terminated as the field approaches the bottom of the 
potential.
Clearly, it is the large initial velocity of the field that results in 
inflation of a finite duration.

\par

In slow roll inflation, the standard Bunch-Davies initial conditions are imposed 
on the perturbations when the modes are well inside the Hubble radius. 
Based on the constraints on the tensor-to-scalar ratio, it is possible to 
arrive at lower bounds on the required duration of inflation (when counted
backwards from its end) if the largest observable scale today is to have 
emerged from sufficiently inside the Hubble radius.
These arguments suggest that (under certain general conditions) inflation 
has to last for {\it at least}\/ $60$--$65$ e-folds in order for all observable
scales to begin their evolution in the sub-Hubble domain (in this context, 
see Refs.~\cite{Dodelson:2003vq,Liddle:2003as}).
In a scenario with kinetically dominated initial conditions, as we mentioned,
inflation naturally lasts for a finite duration.
If this duration is less than the above-mentioned number of e-folds, then a
certain range of large scale modes of cosmological interest would never have
been inside the Hubble radius.
If we now choose to impose the Bunch-Davies initial conditions on the perturbations
(irrespective of whether they are inside or outside the Hubble radius) when
the scalar field rolls down with a large initial velocity, then one finds 
that the scalar power spectrum exhibits suppressed power on the largest 
scales.
Interestingly, if the duration of inflation is chosen suitably, one finds
that the power spectrum improves the fit to the CMB data at the lower
multipoles~\cite{Hergt:2018ksk}.

\par

Typically, features in the inflationary power spectra are generated due to
deviations from slow roll and these departures also lead to larger levels 
of non-Gaussianities (see, for instance, Refs.~\cite{Chen:2006xjb,
Flauger:2010ja,Adshead:2011bw,Adshead:2011jq,Martin:2011sn,Arroja:2012ae,
Hazra:2012yn,Basu:2019jew}).
In this work, we examine if the scalar bispectrum generated in the scenario 
with kinetically dominated initial conditions is consistent with the recent
constraints from Planck on the scalar non-Gaussianity 
parameter~$\fnl$~\cite{Akrami:2019izv}.
We numerically evaluate the scalar bispectrum in such a situation and show 
that, since the initial conditions on the perturbations are imposed at a 
finite early time, the contributions due to the boundary terms in the third 
order action governing the scalar bispectrum dominate over the contributions
due to the bulk terms.
This interesting situation does not seem to have been encountered earlier 
in the literature.
In fact, we find that, for a high initial velocity, the contributions due 
to the boundary terms can enhance the value of the scalar non-Gaussianity
parameter beyond the most recent constraints from Planck.

\par

We shall set $\hbar=c=1$ and $\Mpl=(8\,\pi\,G)^{-1/2}$.
As usual, $a$ and $H$ shall denote the scale factor and the Hubble parameter
associated with the Friedmann universe. 
Moreover, an overdot and overprime shall denote derivatives with respect to 
the cosmic and the conformal time coordinates, respectively.
Further, while $N$ shall denote e-folds, $k$ shall represent the wavenumber 
of the modes.


\vskip 8pt\noindent
\underline{\it Scalar power spectrum in `just enough inflation':}\/~To 
illustrate the suppression of power on large scales that can arise in 
scenarios with kinetically dominated initial conditions, we shall consider 
two models of inflation driven by the canonical scalar field~$\phi$,
\viz the quadratic potential $V(\phi)=m^2\,\phi^2/2$ and the 
Starobinsky model described by the potential $V(\phi)=(\Lambda/8)\,
[1-{\rm e}^{-\sqrt{2/3}\,(\phi/\Mpl)}]^2$.
We start the evolution of the background with a large initial velocity for 
the scalar field such that the first slow roll parameter $\epsilon_1=\dot{\phi}^2
/(2\, H^2 \Mpl^2)$ is initially close to its maximum value, \ie 
$\epsilon_{1{\rm i}}\simeq3$~\cite{Ramirez:2011kk,Ramirez:2012gt,
Handley:2014bqa,Hergt:2018crm,Hergt:2018ksk}.
The initial value of the scalar field is chosen such that inflation lasts for 
about $60$~e-folds. 
The expansion of the universe rapidly slows down the field and inflation sets 
in (\ie $\epsilon_1$ becomes less than unity) after about an e-fold or so.
As the velocity of the field reduces further, it soon settles down on the slow
roll inflationary attractor with a small, nearly constant, velocity.

\par

\begin{figure}[!t]
\begin{center}
\includegraphics[width=8.75cm]{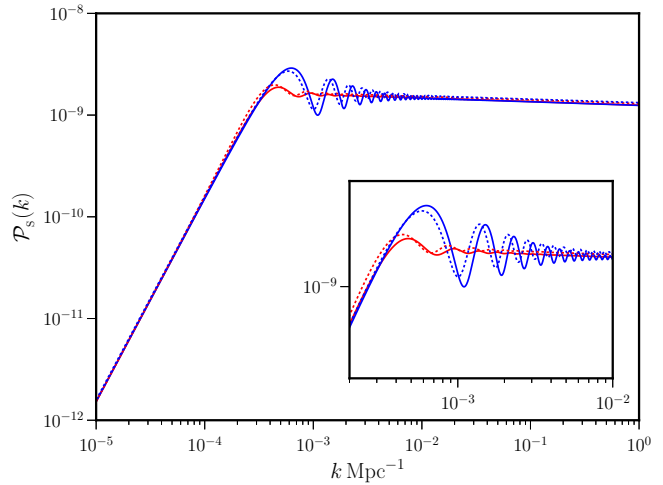}
\end{center}
\vskip -15pt
\caption{The scalar power spectra generated in the quadratic potential (as 
solid lines) and the Starobinsky model (as dotted lines) with kinetically 
dominated initial conditions have been plotted for the two cases wherein 
the perturbations are evolved from $N_{_{\rm I}}$ (in red) and $N=0$ (in 
blue).
We have evolved the field from $\phi_{\rm i} =18.85\, \Mpl$ and $8.3752\,\Mpl$ 
in the quadratic potential and the Starobinsky model respectively, and have set 
$\epsilon_{1{\rm i}}=2.99$.
The parameters~$m$ and $\Lambda$ have been chosen suitably so that the spectra
match over the range of modes which exhibit a suppression in power (for $k<\ki$)
and in the nearly scale invariant regime (which occurs for $k\gg \ki$).
We find that this is possible if we set $m/\Mpl=(5.0\times10^{-6},
4.9\times10^{-6})$ and $\Lambda/\Mpl^4 = (5.8\times10^{-10},5.7\times10^{-10})$
in the cases evolved from $N_{_{\rm I}}$ and $N=0$, respectively.
In these four instances, the pivot scale $k_{\ast}=5\times10^{-2}\,{\rm Mpc}^{-1}$ 
leaves the Hubble radius at $(57.48,58.50)$ and $(57.07, 58.08)$ e-folds 
{\it before the end of inflation}.\/
Also, in these cases, we find that, $\ki/{\rm Mpc}^{-1} = (2.38 \times 10^{-4},
2.32 \times 10^{-3})$ and $(2.38\times10^{-4},1.92\times10^{-3})$.
Note that the two sets of spectra differ only in the amplitude and range
of the oscillations that arise near~$\ki$.}\label{fig:ps}
\end{figure}

We shall numerically evolve the perturbations in such a background and calculate
the resulting observable quantities of interest, \viz the scalar power and
bispectra and the corresponding scalar non-Gaussianity parameter.
Actually, for each model, we shall consider two situations wherein the
perturbations are evolved from two different initial points in time, 
\viz from the onset of inflation (say, $N_{_{\rm I}}$, when $\epsilon_1=1$) 
and from the time (say, $N=0$) when we begin the evolution of the background 
scalar field.
Recall that the Bunch-Davies initial conditions are imposed on the perturbations
when $k \gg \sqrt{z''/z}$, where $z=\sqrt{2\,\epsilon_1}\,\Mpl\,a$.
During slow roll inflation, $\sqrt{z''/z}\simeq \sqrt{2}\,a\,H$, and hence 
the above condition essentially corresponds to the modes being well inside 
the Hubble radius.
Interestingly, we find that the equivalence $\sqrt{z''/z}\simeq a\,H$
proves to be roughly true even when the scalar field is rolling down
the potential with a large initial velocity.
Since we begin the evolution of the perturbations at a specific time, there
naturally arises a finite initial value of the quantity $\sqrt{z''/z}$ 
(evaluated at $N=0$ or $N_{_{\rm I}}$), which we shall refer to as $\ki$.
This implies that, in the scenario of our interest, modes with $k < \ki$ 
would never be inside the Hubble radius.
Despite this, if we were to impose the Bunch-Davies initial conditions on 
{\it all}\/ the modes at the beginning of their evolution (\ie at $N=0$ or
$N_{_{\rm I}}$), then one arrives at a scalar power spectrum with a sharp 
drop in amplitude for modes with $k \lesssim \ki$.
In Fig.~\ref{fig:ps}, we have plotted the scalar power spectra $\ps(k)$ arising 
in these two cases for suitable values of the parameters involved.
Clearly, the spectra exhibit a distinct suppression of power over the modes 
that were never inside the Hubble radius (\ie for $k < \ki$).
The power spectra also contain oscillations (for modes with $k \simeq \ki$)
before they turn nearly scale invariant at smaller scales.
The two sets of spectra presented in the figure differ only in the nature of
the transient oscillations with a higher initial velocity leading to 
oscillations of stronger amplitude and wider range.


\vskip 8pt\noindent
\underline{\it Evaluation of the scalar bispectrum:}\/~The scalar bispectrum 
$G(\vka,\vkb,\vkc)$~---~where $\vka$, $\vkb$ and $\vkc$ constitute a triangular
configuration of wavevectors~---~is determined by the action describing the 
curvature perturbation at the third order (see, for example,
Refs.~\cite{Maldacena:2002vr,Seery:2005wm}).
This action, in turn, is arrived at from the original action governing the 
system of the gravitational and scalar fields. 
The third order action that is often used to calculate the scalar bispectrum 
contains six terms, which are arrived at after repeated integration by parts
(in this context, see, for instance, Refs.~\cite{Martin:2011sn,Arroja:2011yj}). 
In the case of standard slow roll inflation, it can be shown that, barring one 
term, the {\it temporal}\/ boundary terms arising due to integration by parts 
do not contribute to the scalar bispectrum~\cite{Arroja:2011yj}. 
(It can be easily shown that the spatial boundary terms do not contribute in 
any situation.)
The boundary term that contributes (which we shall refer to as the seventh 
term) is often included as a term that arises due to a field
redefinition~\cite{Maldacena:2002vr,Martin:2011sn,Arroja:2011yj}. 
The remaining terms do not contribute in slow roll inflation for two 
reasons:~the contributions from the extreme sub-Hubble domain are 
regulated by the introduction of a cut-off (which is necessary to 
choose the correct perturbative vacuum) and the late time contributions 
prove to be insignificant since the amplitude of the curvature 
perturbation freezes on super-Hubble scales~\cite{Arroja:2011yj}.

\par

\begin{figure}[!t]
\begin{center}
\includegraphics[width=8.75cm]{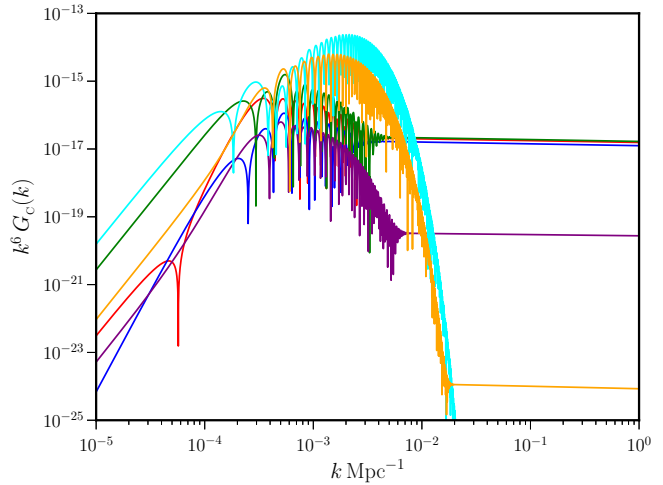}
\end{center}
\caption{The different contributions to the scalar bispectrum in the equilateral
limit~---~the bulk contributions $G_1(k) + G_3(k)$ (in red), $G_2(k)$ (in blue), 
$G_4(k) + G_7(k)$ (in green), $G_5(k) + G_6(k)$ (in purple), and the boundary
contributions $G_8(k)$ (in cyan) and $G_9(k)$ (in orange)~---~have been plotted 
in the scenario wherein the perturbations are evolved from the onset of inflation
in the quadratic potential.
Evidently, the contributions due to the boundary terms dominate the contributions 
due to the bulk terms over a range of modes. 
Moreover, note that the contributions due to the boundary terms prove to be 
considerably more significant around $\ki$, before they die down rapidly on 
smaller scales.
We find that the scalar bispectrum has roughly the same shape in all the models
and cases we have considered.}\label{fig:grrr-1.00}
\end{figure}

In the scenario of our interest, while modes with $k < \ki$ always remain on 
super-Hubble scales, modes with $k>\ki$ begin in the sub-Hubble regime and 
eventually reach super-Hubble scales. 
As the amplitude of all these modes freeze in the super-Hubble regime, the 
contributions due to the boundary terms at late times turn out to be 
insignificant (apart from the seventh term usually taken into account
through a field redefinition) as in the standard slow roll case. 
However, since the modes are evolved from a finite past, we find that we cannot 
ignore the contributions arising due to the boundary terms evaluated at the 
initial time (\ie  at $N_{_{\rm I}}$ or at $N=0$). 
We numerically evaluate the contributions due to the six standard bulk terms
and the seventh term often absorbed through a field redefinition. 
We also calculate the contributions due to all the boundary terms arising at 
the initial time.
In Fig.~\ref{fig:grrr-1.00}, we have illustrated the various contributions 
to the scalar bispectrum in the equilateral limit.
Interestingly, we find that over a range of modes near $\ki$, the boundary 
terms turn out to be comparable to and even larger than the bulk terms.
This is a rather novel result that does not seem to have been encountered
earlier in the literature. 

\par

We had mentioned that the contributions to the scalar bispectrum from the 
sub-Hubble regime are regulated by the introduction of a
cut-off~\cite{Chen:2006xjb,Hazra:2012yn}. 
As is usually done, we introduce a democratic (in the space of wavenumbers) 
cut-off of the form ${\rm e}^{-\kappa\,(k_1+k_2+k_3)/(3\,a\,H)}$, where 
$\kappa$ is a suitable cut-off parameter, when calculating the contributions
due to both the bulk and the boundary terms.
In slow roll inflation or in situations involving brief periods of fast roll
sandwiched between epochs of slow roll, the value of the cut-off parameter 
$\kappa$ is chosen depending on the depth inside the Hubble radius from which 
the integrals characterizing the bulk terms are carried out (in this context, 
see Ref.~\cite{Hazra:2012yn}).
But, in the scenario of our interest, a range of modes (with $k< \ki$) are 
never inside the Hubble radius and another range (with $k> \ki$) do not 
spend an adequate amount of time in the sub-Hubble regime. 
Since the large scale modes (\ie $k < \ki$) always remain on super-Hubble 
scales, the bispectrum evaluated over these range of modes is completely
independent of the choice of the cut-off parameter~$\kappa$.
However, we find that the results depend on the choice of $\kappa$ for modes
around $\ki$ which do not spend an adequate amount of time in the sub-Hubble
regime.
As there exists no definitive procedure that can be adopted to circumvent this 
ambiguity, we make a judicious choice of $\kappa$ for the remaining set of modes 
(\ie for $k>\ki$) based on the natural demand that we are to recover the standard 
slow roll results at suitably small scales which emerge from sufficiently inside 
the Hubble radius (say, $k>10^2\,\ki)$.
We should mention here that we have set $\kappa=0.3$ for {\it all}\/ the modes 
in arriving at the results plotted in Fig.~\ref{fig:grrr-1.00}.
 
\begin{figure}[!t]
\begin{center}
\includegraphics[width=8.75cm]{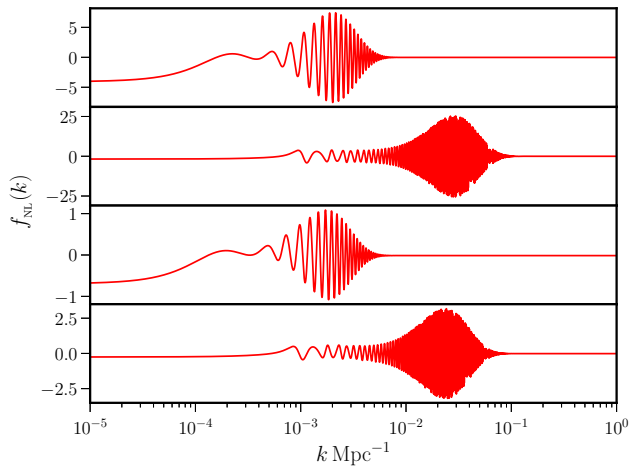}
\end{center}
\caption{The behavior of the non-Gaussianity parameter $\fnl(k)$ has been 
plotted in the equilateral limit for the quadratic potential (in the top 
two panels) and the Starobinsky model (in the bottom two panels) in the 
two cases evaluated from $N_{_{\rm I}}$ (in panels one and three, counted
from the top) and  $N=0$ (in panels two and four).
As expected, the non-Gaussianity parameter exhibits a burst of oscillations 
around $\ki$ before it settles down to the slow roll value at small scales.
In a given model, $\fnl$ is considerably larger in the case wherein the 
bispectrum is calculated from $N=0$ (plotted in panels two and four) than 
in the case wherein it is computed from the onset of inflation (plotted in 
panels one and three).
Interestingly, $\fnl$ is significantly smaller in the Starobinsky model than 
in the quadratic potential.}\label{fig:fnl-eq}
\end{figure}

\vskip 8pt\noindent
\underline{\it Amplitude and shape of $\fnl$:}\/~With the scalar power 
and bispectra at hand, we can now evaluate the resulting non-Gaussianity 
parameter~$\fnl$.
In Fig.~\ref{fig:fnl-eq}, we have plotted $\fnl$ in the equilateral limit 
for the two cases in each of the two models we have considered.
While the quantity has a roughly constant value over wavenumbers $k < \ki$, it 
exhibits oscillations around~$\ki$ before eventually settling down to the 
usual slow roll value for $k\gg \ki$.
Curiously, the constant value at large scales is higher in the case wherein the
perturbations are evolved from the onset of inflation.
However, over the oscillatory regime, the value of $\fnl$ is larger in the case 
wherein the perturbations are evolved from a higher initial velocity of the 
background scalar field.
Importantly, we should clarify that even the largest value of $\fnl$ we encounter 
lies within the constraints (\viz $\fnl=-26 \pm 47$ for the equilateral shape) 
arrived at recently by Planck~\cite{Akrami:2019izv}.
Interestingly, under the same conditions, the amplitude of $\fnl$ turns out to be
significantly smaller in the Starobinsky model than in the quadratic potential.

\par

\begin{figure}[!t]
\begin{center}
\includegraphics[width=8.75cm]{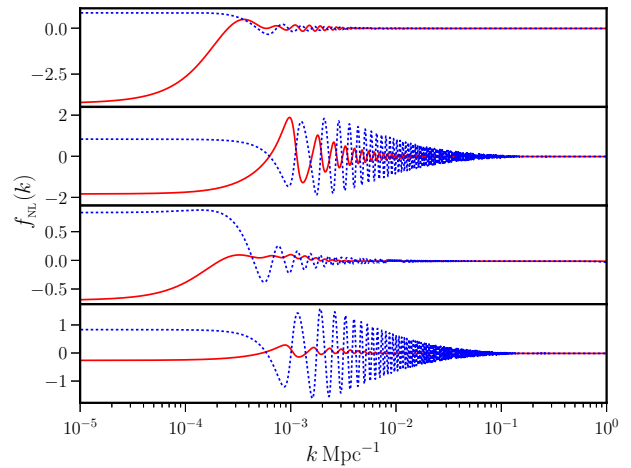}
\end{center}
\caption{The behavior of the non-Gaussianity parameter $\fnl$ in the squeezed 
limit has been plotted for the two models and in the two situations of our
interest as in the previous figure.
We have also plotted the quantity $(5/12)\,(\ns-1)$ (in blue) for all the cases.
It is clear that the consistency condition is violated at large and intermediate
scales before it is restored at suitably small scales (roughly for $k \gtrsim 
10^2\,\ki$).}\label{fig:fnl-sq}
\end{figure}

Let us now turn to the behavior of the scalar non-Gaussianity parameter
in the squeezed limit. 
In Fig.~\ref{fig:fnl-sq}, we have plotted $\fnl$ in the squeezed limit as 
well as the consistency condition, \viz $\fnl^{_{\rm CR}} = (5/12)\,(\ns - 1)$, 
where $\ns$ is the scalar spectral index, again for both the models and in the
two cases of our interest.
Clearly, in the squeezed limit, the non-Gaussianity parameter has broadly the 
same shape as in the equilateral limit. 
It is roughly constant at small wavenumbers, which is followed by a burst of 
oscillations over the intermediate range, before its amplitude is restored to 
the standard slow roll value at larger wavenumbers.
However, the strength of the oscillations in the squeezed limit proves to be
considerably smaller than in the equilateral limit.
Moreover, the consistency condition is violated for the large scale modes (for
which the Bunch-Davies conditions are imposed in the super-Hubble domain), and 
it is eventually restored for small scale modes that emerge from sufficiently 
deep inside the Hubble radius. 


\vskip 8pt\noindent
\underline{\it Conclusions:}\/~The model of `just enough inflation' is attractive
for the reason that it leads to a suppression of power on large scales which 
provides a better fit to the CMB data than a nearly scale invariant spectrum 
produced in conventional slow roll inflation.
It then becomes important to examine whether the non-Gaussianities generated 
in the model are consistent with the recent constraints from 
Planck~\cite{Akrami:2019izv}.

\par

In this work, we have numerically calculated the scalar bispectrum and the 
corresponding non-Gaussianity parameter~$\fnl$ arising in this
scenario for two models, \viz the quadratic potential and the Starobinsky 
model. 
Due to the fact that the initial conditions on the perturbations are imposed in
the finite past when the scalar field is rolling rapidly down the inflationary
potential, the model presents a novel and hitherto unexplored situation as 
far as the calculation of the scalar bispectrum is concerned.
We find that, apart from the standard contributions due to the bulk terms in 
the third order action governing the curvature perturbation, there also arise
contributions to the scalar bispectrum from the temporal boundary terms which 
are usually ignored.
In fact, over a range of modes, we find that the contributions due to the 
boundary terms evaluated at the initial time (when the Bunch-Davies conditions 
are imposed on the perturbations) prove to be dominant when compared to the
contributions due to the bulk terms. 
Moreover, we notice that the extent of the scalar non-Gaussianity generated
depends on the velocity of the scalar field when the initial conditions are
imposed on the perturbations, with the maximum value of $\fnl$ being larger 
when the velocity of the field is higher.
Further, we find that, in the squeezed limit, the consistency condition 
governing the scalar bispectrum is violated for the large scale modes 
which are never inside the Hubble radius and exhibit a suppression in 
the power spectrum. 
Lastly and, importantly, the amplitude of $\fnl$ generated in the quadratic
potential and the Starobinsky model prove to be significantly different. 
These unique signatures of the scenario in the bispectrum can help in 
distinguishing it from other models that achieve similar suppression 
in scalar power and hence may seem degenerate in their performance against
the existing CMB data at the level of the power spectrum. 
We are currently comparing models which lead to a suppressed amplitude on 
large scales in the scalar power spectrum and examining the possibility of 
being able to discriminate them through the bispectrum and its imprints on 
the CMB~\cite{Ragavendra:2019b}.


\vskip 8pt\noindent
\underline{\it Acknowledgements:}\/~The authors wish to thank Dhiraj Hazra for
discussions and comments on the manuscript.
HVR would like to thank the Indian Institute of Technology Madras, Chennai, 
India, for financial support through half-time research assistantship.
DC would like to thank the Tata Institute of Fundamental Research, Mumbai, 
India, for financial support. 
HVR and LS wish to acknowledge the use of the cluster computing facility at the
Department of Physics, Indian Institute of Technology Madras, Chennai, India, 
where some of the numerical computations were carried out.

\bibliographystyle{apsrev4-2}
\bibliography{sbs-in-kdi}

\end{document}